\documentclass[aps,prl,twocolumn,groupedaddress,showpacs,floatfix]{revtex4}
\usepackage{graphics}
\usepackage{epsfig}
\begin{document}
\title{Origin of complex crystal structures of elements at pressure}
\author{G.J.Ackland, I.R.Macleod and O.Degtyareva}
\affiliation{School of Physics and Centre for Science at 
Extreme 
Conditions, The University of Edinburgh, Mayfield Road, Edinburgh, EH9 3JZ, UK.}

\begin{abstract}

We present a unifying theory for the observed complex structures of 
the sp-bonded elements under pressure based on nearly free electron 
picture (NFE).  In the intermediate 
pressure regime the dominant contribution to crystal structure arises from 
Fermi-surface Brillouin zone (FSBZ) interactions - structures which allow this are favoured.    
This simple theory 
explains the observed crystal structures, transport properties,
the evolution of internal 
and unit cell parameters with pressure.  We illustrate it with 
experimental data for these elements
and ab initio calculation for Li.

\end{abstract}
\date{\today}
\pacs{61.50.Ks,62.50.+p}    
\maketitle

\begin{figure}[ht]
\protect{\includegraphics[width=\columnwidth]{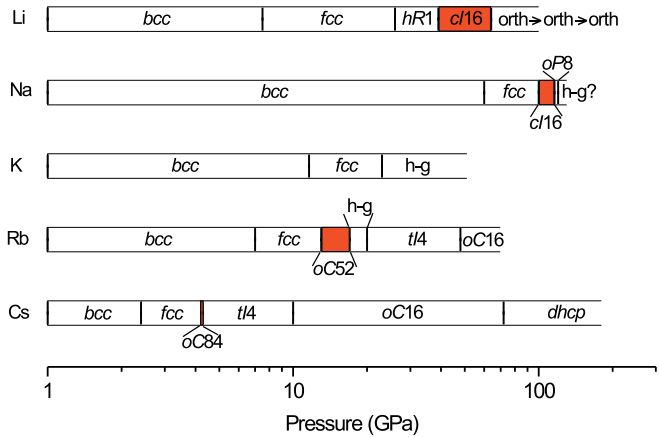}}
\caption{High-pressure structural transition sequences for the alkali
and alkaline earth elements\cite{Li-cI16,Na,K,K-III,Rb-III,Cs-III,
Rb-IV,Cs-IV,Rb-V,Rb-VI,Cs-dhcp}. 
The general pattern is similar, with low-Z materials having 
higher transition pressures. Intermediate structures are labelled in Pearson notation. 
\protect\label{fig:structure}}
\end{figure}

Recent experimental high-pressure investigations of metallic elements
have yielded surprising and intriguing results, with 
decreasing coordination and increasing crystal complexity at 
intermediate pressures (fig.\ref{fig:structure}).
This behavior can be
reproduced by ab initio calculations based on the density
functional plane wave pseudopotential method (DFPP) which represents a
sufficient theoretical understanding of the problem: there is no need
for physics ``beyond'' DFPP, as is required in the $f$-metal 
localisation transitions.
However, DFPP calculations are
sufficiently complex that no simpler theoretical principles emerge.
Moreover DFPP calculations require plausible candidate structures
- in simple cases picking the ``usual suspects'' and symmetry-breaking
distortions therefrom has worked, but when very complex structures 
are contenders, a more systematic approach is needed.  Here, we
demonstrate general principles which contribute to complex
crystal structures, and deduce heuristics from which to choose candidate
structures.

DFPP gives the enthalpy of various structures,
the lowest enthalpy structure being stable. 
Under pressure the total enthalpy comprises a band structure 
term (the eigenvalues of 
non-interacting electrons moving in an effective field), an Ewald 
sum, pressure times volume, 
exchange correlation and Hartree energies.\cite{KS} 
The dependence on volume ($\Omega$) is as follows:
$\Omega^{-1/3}$ for the coulombic terms (Hartree, ion-ion, ion-electron), 
$\Omega^0$ for the exchange correlation (neglecting 
``non-linear core corrections''), $\Omega^{-2/3}$ for kinetic energy.
Thus under pressure materials become more free-electron like.
We note two aspects of DFPP.  Firstly, the fact that pseudopotentials work 
indicates that repulsion between core electrons can be neglected.  
Secondly, energy minimization in DFPP codes  
is hugely improved by {\it preconditioning}\cite{mcp}: assuming that the
contribution from a plane wave basis state is primarily its free-electron-like
kinetic energy.

Two further effects are 
not explicit here: imperfect screening of ionic charge as ions 
approach closely and the FSBZ energy splitting from interaction between plane waves and ionic potentials\cite{MottJones}: 
$\Delta E = \pm \int_\Omega e^{i{\bf k.r}} V({\bf q.r})  e^{i{\bf k'.r}} d^3{\bf r}
\;.$

V({\bf q.r}) depends on the scattering at {\bf q=k+k'} which, for elements,
is proportional to the X-ray diffraction.  If {\bf k} falls near the Fermi 
level the state with increased energy is unoccupied, while the other 
is occupied. Thus FSBZ interaction gives a first order 
change in energy with crystal structure, while at 
other {\bf k} the energy gained and lost cancels.
FSBZ effects scale as $\Omega^{-1}$.  

It is important to distinguish those terms which contribute most 
to the {\it total energy} from those which contribute to 
{\it energy differences between crystal structures}.
The latter
are the screened ion-ion potential and the perturbation of the free electrons 
from FSBZ interactions.  Given its  $\Omega^{-1}$ scaling with volume, 
FSBZ interaction may dominate at intermediate pressures with the
$exp(-\Omega^{1/3})$ dependence of imperfect screening being important at the 
highest pressures.

The central result of this paper will be the demonstration that the
complex elemental crystal structures at intermediate pressures 
can be simply understood using NFE.
This theory also gives a way of picking plausible trial structures 
for total energy calculation, and explains a number of observed properties.
To ease comparison with experiment, we will refer to specific 
interactions between the FS and particular points in the BZ 
(diffraction peaks) rather than the equivalent\cite{planepoint} description in terms of 
BZ planes containing these points.

\begin{figure}[ht]
\protect{\includegraphics[width=\columnwidth]{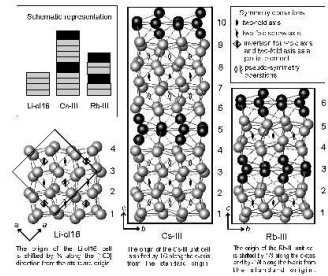}}
\protect{\caption{Crystal structures of 
Li-cI16(left)  Cs-III (middle) and Rb-III (right)
viewed along the [100] axis, 
of the $C222_1$ unit cell. 
Interatomic distances up to 2.5{\AA}(Li), 4.7{\AA}
(Cs) and 4.1{\AA}(Rb) are shown as solid lines. True symmetry 
operators normal to the plane of projection are shown 
by solid symbols,
pseudo-symmetry operators by open symbols.
The $cI$16 structure can be viewed as a simple distortion of
body-centered cubic structure, in which each atom is
shifted by $x$ along the bcc
[111] directions \cite{Li-cI16}:
in this projection we find  8-atom layers. 
For Cs-III and Rb-III\cite{Rb-III} identical 
8-atom layers exist\cite{layerdetail} interspersed with 
10-atom layers (black) which can be considered as 8-atom layers with
(001) dumbbell interstitials inserted. 
Thus, the Cs-$oC$84 and Rb-$oC$52  structures are the same as
Li-$cI$16 with density increased by interstitial atoms every 
fifth (Cs) or third (Rb) layer.
\label{layers}}}
\end{figure}

\begin{figure}[ht]
\protect{\includegraphics[width=0.9\columnwidth]{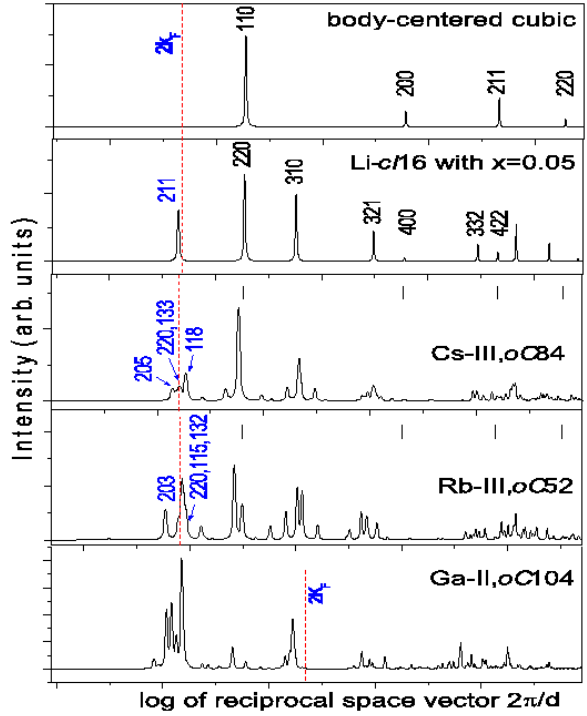}}
\caption{Calculated diffraction patterns for bcc 
the complex structures of Li, Cs, Rb and Ga.  The major 
effect of the  $cI$16 distortion in Li is to throw up a 
211 peak just below the Fermi Vector\protect{\cite{bigger}}, calculated assuming a 
free electron sphere to equate to $2\theta=2\sin^{-1}(\lambda 
\sqrt[3]{3N/64\pi V})$
\label{fig:FSBZ}}

\end{figure}

Under pressure, group I and II metals transform from simple 
structures, bcc and fcc, to complex structures.
The striking similarity 
between the Cs-$oC$84, Rb-$oC$52  and  
Li-$cI$16 structures\cite{Li-cI16,Cs-III,Rb-III} is shown in Fig.\ref{layers} 
and in the diffraction patterns (Fig.\ref{fig:FSBZ}): 
all are derivatives of bcc.

\begin{figure}[ht]
\includegraphics[width=\columnwidth]{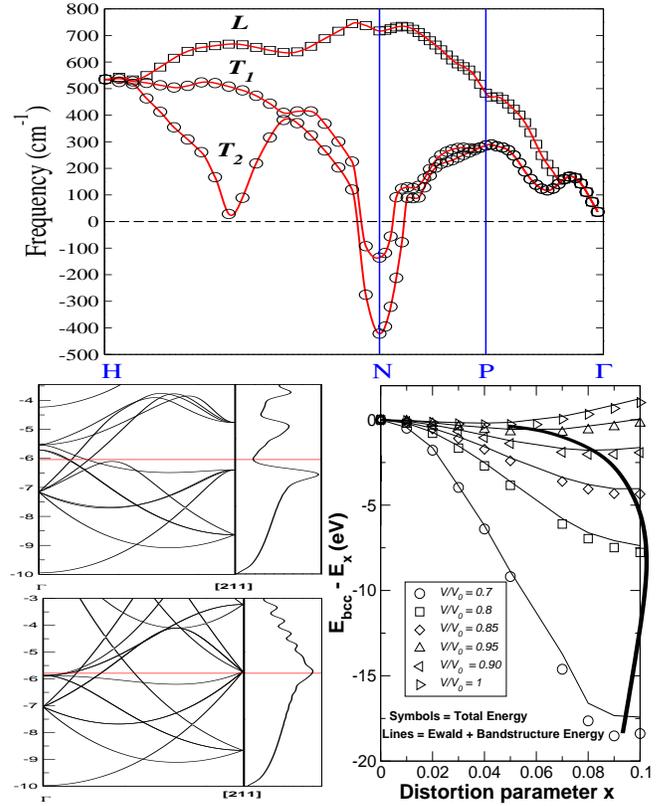}
\caption{ Top: Ab initio calculation of phonon dispersion 
relation\cite{phon} showing dynamic instability for bcc-Li at 40GPa. 
Bottom: (right) Li-DFPP energy difference between bcc and cI16 for various x 
over a range of volumes ($V_0$ =$146.4A^3$, the volume at which cI16 is 
first observed). 
The structure is stable between 42-50 GPa  V/$V_0$=1-0.85. 
Symbols are total energy, lines are Ewald and bandstructure only, 
showing that the Hartree and exchange-correlation contributions 
are negligible. The thick line indicates the optimum value of x at each volume.
(left) Band structures along the [2x,x,x] direction for $V_0$, 
x=0 (lower panel) and x=0.05 (upper panel) 
showing the opening of a pseudogap due to FSBZ interactions: the
surface touches the 24 (211) BZ planes.  
Elsewhere in the BZ  bcc and cI16 band structures are similar.
\label{fig:Evx}}
\end{figure}

Complex phase stability in Rb and Cs\cite{Cs-III,Rb-III} has been attributed 
to $s-d$ transfer of electrons allowing directional bonding and hence 
open structures.  A strong justification for this was the now-discredited 
isostructural phase transition in Cs\cite{Cs-III} which cannot be explained in a free-electron picture.  In a localised basis picture $s-d$ transfer 
transfer certainly exists, but offers no insight into the nature of 
the structures and cannot explain the Li-$cI$16 structure(Fig.\ref{layers}) 
as there are no d-orbitals available to Li.
DFPP calculations for 
$cI$16 Li  provide an ideal testbed for the alternate FSBZ picture
(Fig.\ref{fig:Evx}):
FSBZ effects  scale strongly $\Omega^{-1}$ with reduced volume, 
causing increases in internal parameter $x$, 
(211) band gap opening and diffraction peak intensity. 
At very high pressure this is overcome by
imperfect screening of ionic charges (exponential in 
$x\Omega^{1/3}$).

We have calculated the optimal value of $x$ in $cI$16 Li using DFPP.
 Figure.\ref{fig:Evx} shows variation of total energy with 
atomic position parameter x, for a range of volumes.  
Ewald and band structure contributions account 
for all the variation:  for larger x the band 
structure energy is lowered, but the Ewald energy also 
increases and ultimately becomes dominant.
Thus, the band structure energy 
is responsible for complex structure stability in Li and
\cite{Pucci97,Li-cI16,NT99,NT01}. 
the signature in powder diffraction is that 
one or more diffraction peaks approach $k_{F}$\cite{bigger}.

Precisely this signature 
exists in Cs and Rb Figure\ref{fig:FSBZ}, indicating similar physics.  
The interstitial atoms reduce volume per atom at the expense of splitting the
the diffraction peak near to the Fermi level.  In 
a real space picture the sequence of first increasing then decreasing 
interstitial number seems inexplicable.  In the FSBZ picture, the Fermi 
surface is more easily deformed at large Z.  For Li, the PV enthalpy 
gained from interstitials cannot compensate for loss of FSBZ 
interaction from peak splitting. Rb has FSBZ interactions
with three of the four split peaks, while Cs deforms to touch all four.

This gives us the desired heuristic in the search for candidate ``complex'' structures 
for DFPP calculation: multiple strong diffraction peaks at the Fermi energy.
It also suggests that a simple model of interatomic interactions in such 
materials should concentrate on a particular wavenumber in reciprocal 
space: a condition somewhat encapsulated long ago by Friedel 
oscillations.
 

With the FSBZ picture established for alkali metals, it is instructive 
to turn to other elements.  
Recently, complex Li-type distorted bcc 
structures have been reported in gallium\cite{Ga-II}. 
This extraordinary confluence of monovalent and trivalent materials would 
appear to contradict the FSBZ picture.  However, fig.\ref{fig:FSBZ}
reveals the (310)$_Li$-type diffraction peak just  
below the Fermi level.  Such effects 
have also been observed in intermediate pressure phases of Group IV and 
III-V compounds\cite{rev}.

In Rb, Ba and Sr
a new principle emerges: In real space 
BaIV-type forms self-hosting ``hotel'' structures which can only 
be described using two interpenetrating crystal structures 
(the ``host'' and chains of ``guests'') \cite{skr,Ba-IV,Sr-V,Rb-IV}.  The 
structures can be incommensurate, giving rise to two distinct Brillouin 
zones and the consequent additional possibilities for FSBZ interactions.
Again their stabilities are well reproduced by DFPP\cite{skr}.  

\begin{figure}[ht]
\protect{\includegraphics[width=0.9\columnwidth]{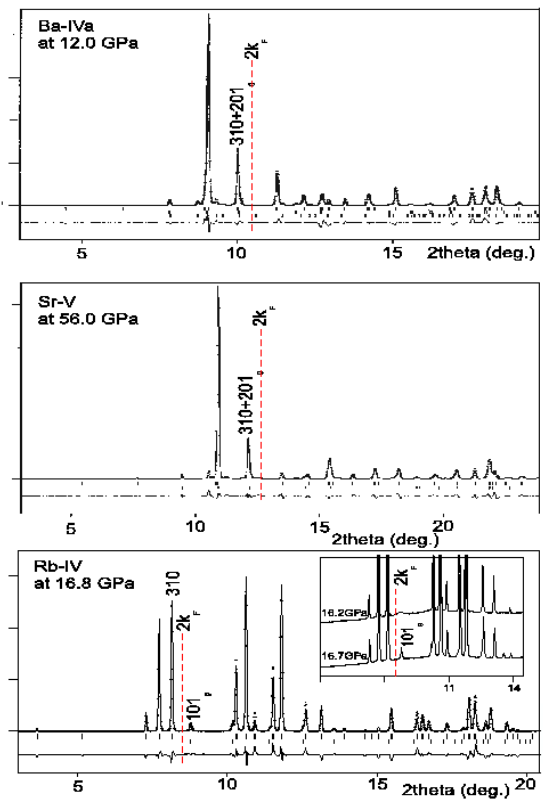}}
\protect{\caption{Calculated diffraction patterns for
hotel structures of Ba-IV, Sr-V and Rb-IV.  The tick marks show 
whether the reflection is from host (upper) or guest (lower) lattice. 
The insets show how the Brilloiun zone maintains contact with the 
Fermi surface.
}\label{fig:hotel}}
\vspace{-2mm}
\end{figure}

Figure \ref{fig:hotel} shows their 
diffraction patterns - once again strong diffraction peaks lie 
near the Fermi level, and the FSBZ picture is dominant.  Significantly, 
while in Ba and Sr there is interaction with the (201) guest, there 
is no guest reflection (with non-zero $k_z$) near the BZ in Rb.  Thus 
FSBZ interaction  order the positions of 
adjacent chains in Ba and Sr, but not in Rb:  Indeed, interchain order is
observed in Ba and Sr, but Rb undergoes a ``melting'' transition at 
low pressure.  Similarly, the ratio between guest and host lattice 
parameters is pressure independent in Ba and Sr, (locked by the FS-$(201)_g$ 
interaction), but pressure dependent in Rb, since no FSBZ effects fix the 
guest c/a ratio.

Other properties of the complex crystal structures in alkali metals 
relate to the FSBZ interpretation. 

The resistivity in Li\cite{Li-resist}, rises tenfold between
40-120 GPa,  corresponding to the $cI$16 and  
other complex orthorhombic phases \cite{Li-cI16, Na}. 
Similar behavior is observed in Cs
(4 GPa, the Cs-III phase) and Rb (10GPa, the Rb-II and Rb-III phases)
\cite{Cs-Rb-resist}. 
Fig.\ref{fig:Evx} shows that
the FSBZ interaction which open 
pseudogaps at the Fermi level in 
Li-$cI$16 giving much lower
electron density at the Fermi level than in fcc
and bcc phases. 

Superconductivity arises from coupling of electrons to low frequency 
vibrational modes; 
we have not performed detailed calculations, but
complex phases tend to exhibit superconductivity through the 
the low frequency phonons associated with FSBZ effects, and 
phasons of incommensurate phases. 
Superconductivity is observed in complex structures of 
Li $cI$16 phase\cite{Li-cI16},
Cs, Ba, Sr and Ga.
In contrast superconductivity is not detected 
in Rb up to pressures of 21 GPa\cite{LiCsGaRb-supercond}, where 
the weak non-FSBZ interchain coupling melts the phason mode.


Projection of our wavefunctions onto atomic orbitals 
shows increased $d$ character with pressure, however, 
we have shown that the concept of FSBZ interactions provides a better
simple description for the ``complex'' phases observed under pressure than 
does $s-d$ transfer.  
A simple heuristic for stable phases is the existence of strong diffraction 
peaks just below the free-electron Fermi vector\cite{bigger}.  Monovalent 
elements achieve this by distortion of bcc with FSBZ at (211) peaks;  
divalent materials form host-guest structures with two Brillouin Zones 
and trivalent gallium adopts distorted-bcc structures with FSBZ at 
(310) peaks.  Unlike s-d transfer, the theory also accounts 
qualitatively for the chain-melting and pressure sensitivity of c/a in 
rubidium, pressure independence of c/a ratio in Ba and Sr, the increased
superconductivity transition temperature in complex phases, 
and the similarity of Li with other group-1 elements under pressure.

We acknowledge discussions with M.I.MacMahon, R.J.Nelmes and V.Degtyareva.  
Calculations were done using the VASP and PWscf packages\cite{packages}


\begin{thebibliography}{99}
\bibitem{planepoint} Strictly, FSBZ energy from second order perturbation 
theory requires an integral between pairs of occupied and unoccupied 
orbitals in adjacent BZs.  This integral is complicated enough to 
obscure the simple picture, yet quantitatively less accurate than DFPP
calculation - since its value varies with the interaction at the 
diffraction peak, we use the latter for qualitative discussion.
\bibitem{layerdetail}Layers are 
inclined at about $\pm35^o$ to the conventional $b-$axis \cite{Rb-III}. 
The angle of $35^o$ originates  as the angle between bcc planes
(110) and (111); it does not change in the bcc-CI16 transformation.
\bibitem{bigger} In 
DFPP some high-k components are mixed into low energy 
states, so the free-electron-like states at $E_F$ 
will have wavevectors below the free electron $k_F$.  
For maximum BZFS effect $k_F$ should lie just above 
the relevant diffraction peak. $s-d$ transfer arguments 
give a similar picture by assuming fewer free electrons.
\end{thebibliography}
\end{document}